\documentclass[sigconf]{acmart}

\AtBeginDocument{%
  \providecommand\BibTeX{{%
    \normalfont B\kern-0.5em{\scshape i\kern-0.25em b}\kern-0.8em\TeX}}}

\setcopyright{rightsretained} 
\acmConference[EARS'19]{SIGIR 2019 Workshop on ExplainAble Recommendation and Search}{July 25, 2019}{Paris, France}

\usepackage{booktabs} 
\usepackage{url}
\usepackage{algorithm}
\usepackage{amsmath,amsthm}
\usepackage{algpseudocode}
\usepackage{amssymb}

\DeclareMathOperator*{\argmax}{arg\,max}
\usepackage{graphicx}
\newcommand{\sysname}[1]{\textsc{#1}}

\begin{document}

\title[Hierarchical Item Categories for Efficient Recommendations and Browsing]{Learning Hierarchical Item Categories from Implicit Feedback Data for Efficient Recommendations and Browsing}



\author{Farhan Khawar}
\affiliation{
  \institution{Department of Computer Science \& Engineering\\The Hong Kong University of Science and Technology}
  \city{Hong Kong}
  \country{Peoples Republic of China}
}
\email{fkhawar@cse.ust.hk}

\author{Nevin L. Zhang}
\affiliation{
  \institution{Department of Computer Science \& Engineering\\The Hong Kong University of Science and Technology}
  \city{Hong Kong}
  \country{Peoples Republic of China}
}
\email{lzhang@cse.ust.hk}


\begin{abstract}
Searching, browsing and recommendations are common ways in which the ``choice overload'' faced by users in the online marketplace\footnote{We use the generic term marketplace here to refer to online vendors, portals, content providers etc. } can be mitigated. In this paper we propose the use of hierarchical item categories, obtained from implicit feedback data, to enable efficient \emph{browsing} and \emph{recommendations}. We present a method of creating hierarchical item categories from implicit feedback data only i.e., without any other information for the items like name, genre etc. Categories created in this fashion are based on users' co-consumption of items. Thus, they can be more useful for users in finding interesting and relevant items while they are browsing through the hierarchy. We also show that this item hierarchy can be useful in making category based recommendations, which makes the recommendations more explainable and increases the diversity of the recommendations without compromising much on the accuracy. Item hierarchy can also be useful in the creation of an automatic item taxonomy skeleton by bypassing manual labeling and annotation. This can especially be useful for small vendors. Our data driven hierarchical categories are based on hierarchical latent tree analysis (HLTA) which has been previously used for text analysis. We present a scaled up learning algorithm \emph{HLTA-Forest} so that HLTA can be applied to implicit feedback data. 
\end{abstract}



\maketitle

\section{Introduction}

The online marketplace offers virtually an unlimited shelf-space for the items that are available for the users to consume. This results in choice-overload \cite{Bollen:2010:UCO:1864708.1864724} and unwanted distractions that make it difficult for the users to find the items they desire. Three common ways to tackle this issue are letting users: search the item space via specific queries, browse the item space manually and consume the recommendations provided. In this paper we show how learning hierarchical item categories from implicit feedback can be used for efficient browsing and recommendations.

Implicit feedback \cite{nichols1997implicit} is arguably the simplest and easiest form of user feedback to collect. It only contains the information of which items a user has consumed before, and no information about what users are not
interested in \cite{Pan:2008:OCF:1510528.1511402}. Yet user preferences and behaviors can be elicited effectively from this seemingly simple feedback since it captures the users' natural behavior \footnote{It is not affected by ratings biases found in explicit feedback e.g. users not rating items they don't like or users not revealing their true preference on controversial/taboo items etc.}.

An  important information present in implicit feedback is the information of co-occurrence of items i.e., which items are co-consumed by users. Using this key information we propose to use HLTA to learn item hierarchies. When applied to implicit feedback data, HLTA can obtain item category hierarchy by first identifying item co-consumption patterns
(groups of items that tend to be consumed by the same users, not necessarily at the same
time), then identifying co-occurrence patterns of those patterns, and so on. Therefore, the lowest level is a group of items and they form the most specific categories, and the next level has more general categories which are groups of these specific categories and so forth.

Items are usually organized hierarchically in categories. This is because most items, especially products in online stores, naturally group in hierarchical categories
. However, creating and maintaining these hierarchies is a tedious job. They require manual annotation, labeling and/or description of the items. Such information, although can be obtained, is not always easily available \footnote{Product descriptions are generally available, but since they are manually specified, the quality of information is subjective. Therefore, hierarchies created based on such information are also affected by this noise.}. This is especially true for small vendors who have limited resources. Since implicit feedback is very easy to collect and requires no manual intervention, creating automatic item hierarchies from implicit feedback is very attractive.

Once such hierarchy is available, the users can browse the item space efficiently. They can first choose the broadest category they are interested in and then zoom in the hierarchy by selecting a sub-category and so on until they reach their desired items. This allows the users to explore the item space for new interests, or to find items they know exist but don't recall their specific details to search for them, or to simply stumble upon items they find interesting. Browsing the item space in this manner requires \emph{active} participation by the user.

Recommending items to users does not require the users to actively explore the item space. Users can simply receive recommendations while normally interacting with the system. In this setting we can also use the item categories to make recommendations. We can see which categories the user has consumed the most items from previously and recommend new items to the users based on the proportion of items the user consumed from each of these categories. The recommendations within each category can be made based on any existing recommender. We dub such a recommender as a ``Category Aware'' recommender (CAR).

We will see later CAR results in recommendations that are more diverse and explainable and with roughly the same accuracy as the base recommender. Both diversity, measured as the number of unique items recommended by the recommender, and explanability, the ability of the recommender to provide explanations for the provided recommendations, are desirable properties of a recommender. And CAR is able to provide both of these improvements to a base recommender \emph{simultaneously}. The main contributions of this paper are as follows:
\begin{itemize}
	\item Presenting a scalable and memory efficient algorithm for learning hierarchical item categories from implicit feedback only.
	\item Recognizing that item hierarchies can be useful in mitigating choice overload by enabling category wise browsing. And providing an example of such a system. 
	\item Showing that the item categories can be used to make diverse and personalized recommendations given any base recommender.
	\item Showing that the item hierarchies also allow us to make a base recommender explainable. 
\end{itemize}

\section{Hierarchical Latent Tree Model}\label{hltms}

\begin{figure*}
\begin{center}
\includegraphics[width=15cm]{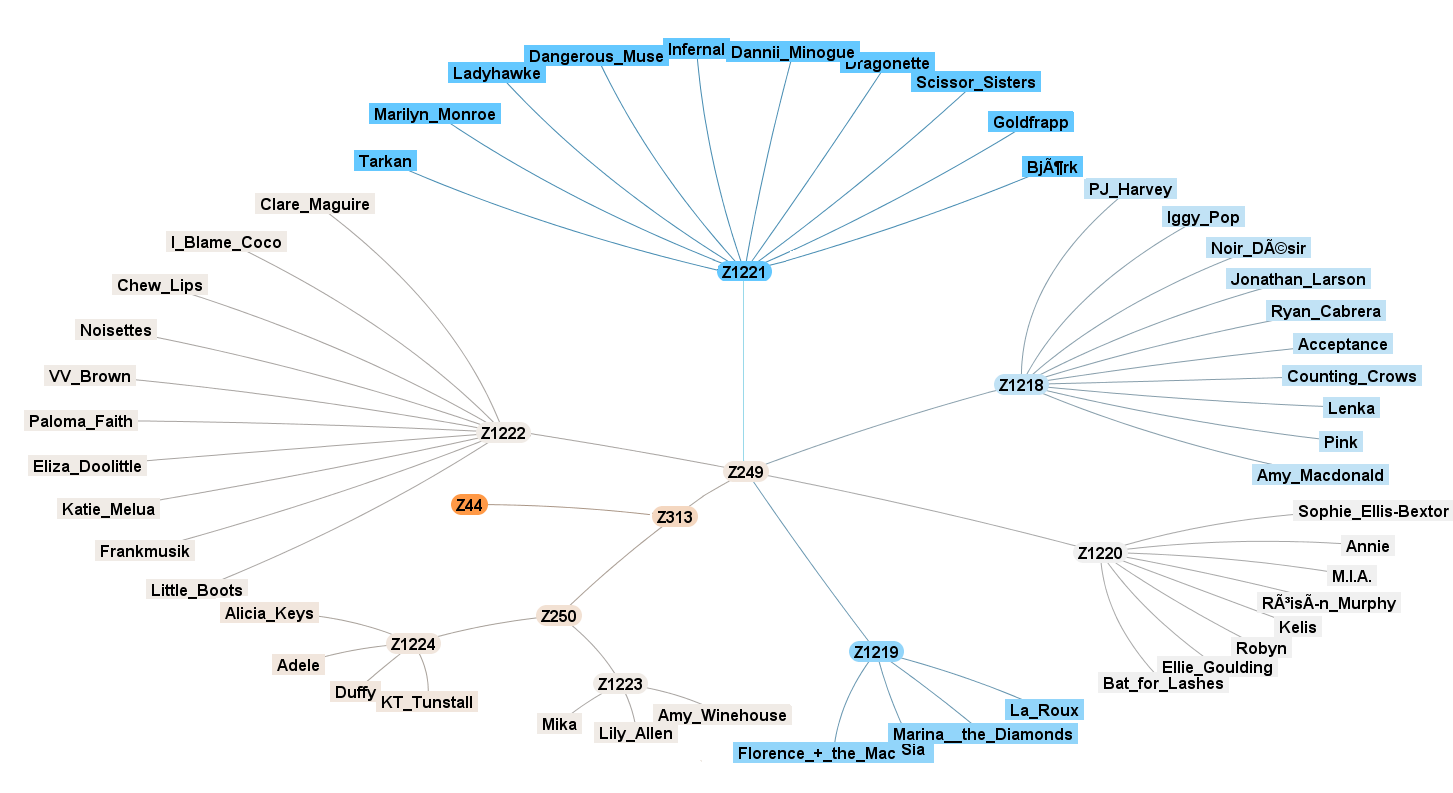}
\caption{Part of a hierarchical subtree rooted at the latent variable $Z_{313}$ which primarily represents the pop-rock category and some portion of pop-hiphop. The observed variables are the items (artists) that the users have listened to. The latent variables start with the prefix ``$Z$'' and represent a category of items. The fist digit after ``$Z$'' denotes the level of the latent variables e.g. $Z_{44}$ is a fourth level latent variable, $Z_{313}$ is a third level latent variable and $Z_{1220}$ is a first level latent variable. }
\label{fig:z313}
\end{center}
\end{figure*}

Hierarchical Latent Tree models (HLTMs) are tree structured models with a layer of
observed variables at the bottom, and multiple layers of latent variables on top \cite{chen2017latent,khawar2019conformative,khawar2018using}.  When applied to implicit feedback data, HLTA learns models such as the one shown in Figure \ref{fig:z313}, which is a part of the model that was obtained from the last.fm dataset\footnote{\url{https://grouplens.org/datasets/hetrec-2011/}.\\ We use this dataset for illustration purposes since artist names are easily interpretable by humans.}.

In this paper, all the variables are assumed to be binary. The observed variables represent items and indicate whether the items were consumed by a user. For example, the value of the variable {\tt Duffy} for a user is 1 if she listened to {\tt Duffy} before, and 0 otherwise. The latent variables are introduced during data analysis to explain co-consumption patterns detected in data.

In the model in Figure \ref{fig:z313}, the artist variables are partitioned into disjoint clusters. Each disjoint cluster is a group of artists that tend to be co-consumed, in the sense that users who listened to one of them often also listened the others. There is a latent variable for each cluster of artists. This represents a category of artists e.g. $Z_{313}$ represents pop artists. And the latent variables are connected up to form a tree structure. Each edge in the tree is associated with a conditional probability distribution.

We see that artist category $Z_{250}$  is a child of the category $Z313$. The first child of $Z_{250}$ is $Z_{1224}$ which represents the artists {\tt Alicia Keys}, {\tt Adele}, {\tt Duffy} and {\tt KT Tunstall}. All of these artists are female pop-hiphop and pop-rock singer-songwriters. Therefore, $Z_{1224}$ represents the category ``female pop-hiphop/pop-rock singer-songwriters''. The second child of $Z_{250}$ is $Z_{1223}$ and all of the artists under $Z_{1223}$ are "English pop-soul-rock singer-songwriters". Therefore, $Z_{250}$ defines the category of ``mostly female singer-songwriters with a flavor of pop and pop-rock''. 

Another child of $Z_{313}$ is $Z_{249}$ which has five more children. Of them, $Z_{1222}$ represents British electronic-pop artists,  $Z_{1218}$ represents pop-rock/punk artists that are primarily non-British, $Z_{1220}$ are female vocalists who sing or play electropop-R\& B-dance etc. But what puts all of them in the same category is that they represent artists that fuse pop with other genres of music
. Therefore, $Z_{249}$ represents pop singers who have another dimension, like electronic, hip-hop, rock, indie etc., to their songs too. We see that the genres of artists in $Z_{249}$ and $Z_{250}$ are related but artists within each category are more related than artists between categories. Also, it is not easy to succinctly describe the genres of artists under each category by words or labels, however, it is much easier to see their relationships from a category tree that has been learned based on co-consumptions. Similarly, other latent variables specify other categories of artists. The latent variables that are closer to the leaves of the tree represent more specific categories and as we go higher the categories become more general.

\section{Learning HLTM}\label{hltm}
HLTA was originally proposed for text analysis where the number of observed variables are not of the order of items in a recommender system, therefore, a straight forward application of HLTA to implicit feedback is not practical. In this section we go over the key points of learning an HLTM and present our algorithm HLTA-Forest that can scale up for implicit feedback
. An overview of our learning algorithm HLTA-Forest for implicit feedback data is given in Algorithm \ref{alg.top}.

\begin{algorithm}[t]
\begin{description}
\item[Inputs:] $\mathcal{R}$ --- a collection of binary user histories, $\tau$ --- upper bound on the number of top-level categories, $\mu$ --- maximum category size, $\delta$ --- threshold used in model building test, $\kappa$ --- number of EM steps for each category. \\
\item[Outputs:] An HLTM $m$.
\end{description}

\begin{algorithmic}[1]
\State $\mathcal{R}_1 \gets \mathcal{R}$, $m \gets null$.
\Repeat \label{alg1.start}
    \State $m_1 \gets$ \sysname{LearnFlatForestModel}($\mathcal{R}_1$, $\delta$, $\mu$, $\kappa$);\label{alg1:learnFlat}
    \State $\mathcal{R}_1 \gets$ \sysname{HardAssignment}($m_1$, $\mathcal{R}_1$); \label{alg1.hardassign}

    \If{${m}= null$}
        \State ${m} \gets {m}_1$;\label{alg1.m}
    \Else
        \State $m\gets$ \sysname{StackModels}(${m}_1$, ${m}$); \label{alg1.stackmodel}
    \EndIf

\Until number of top-level nodes in $m  \leq \tau$.  \label{alg1.end}
\State Link the top level latent nodes of $m$ to form a tree.
\State \Return $m$. \label{alg1.return}
\end{algorithmic}
\caption{\sysname{HLTA-Forest}($\mathcal{R}$, $\tau$, $\mu$, $\delta$)}
\label{alg.top}
\end{algorithm}

\begin{figure}

\begin{center}
\includegraphics[scale=0.25]{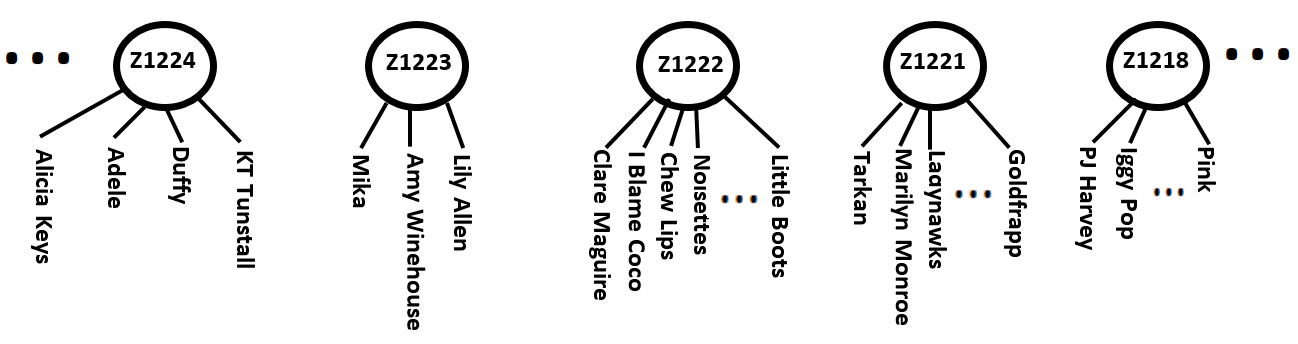}
\caption{Part of a Level-1 flat model learned by HLTA-Forest for the last.fm dataset. Each latent variable is connected to at least three observed variables forming a tree. Together all these trees form a forest. }
\label{fig:forest}
\end{center}
\end{figure}

The procedure in Algorithm \ref{alg.top} has three key steps: Learning a flat model (\sysname{LearnFlatForestModel}), hard-assignment ( \sysname{HardAssignment}) and stacking the flat models (\sysname{StackModels}). Given the implicit feedback user histories $\mathcal{R}$, the first step (line~\ref{alg1:learnFlat}) results in a flat model in which each latent variable is connected to at least one observed variable. This flat model is a forest with each latent variable having its own tree e.g. the level-1 latent variables $Z_{1xxx}$ in Figure \ref{fig:forest} each have their own tree and together they form a forest. The learning of a flat model is the key step of HLTA.

The second step (line~\ref{alg1.hardassign}) assigns each user to a particular state of each latent variable. This effectively makes the latent variable to behave like an observed variable for model creation purposes. This is done by carrying out inference on the learned model $m_1$ to compute the posterior distribution of each latent variable for each user. The user is assigned to the state with the highest posterior probability, resulting in a dataset $\mathcal{R}_1$ over the level-1 latent variables. Note that in our procedure inference is carried on $m_1$ using $R_1$ unlike \cite{chen2017latent,khawar2018using} where inference is done on $m$ using $R$. More details on our \sysname{HardAssignment} routine are provided later.

Next we execute line~\ref{alg1:learnFlat} again to learn another flat latent tree model (LTM), this time for the level-1 latent variables. In the third step (line~\ref{alg1.stackmodel}), we stack the flat model for the level-1 latent variables on top of the flat model for the observed variables, resulting in the hierarchical model. The parameter values for the new model are copied from the two flat  models. In general, the first three steps are repeated  until the number of top-level latent variables falls below a user-specified upper bound  $\tau$ (lines~\ref{alg1.start} to~\ref{alg1.end}). 

 It is important to note that this general procedure is similar but not identical to the HLTA learning algorithm for text analysis. Specifically, we do hard assignment of users to latent variables after each flat model. Moreover, since we are interested in the HLTM structure we do not run EM on the whole model $m$. And, finally we only link the top level latent nodes to form a tree. All intermediate flat models are forests. Next we describe the specific details.

\subsection{Learning Flat Model}
In this section we describe our learning flat model routine that can scale up HLTA for collaborative filtering datasets. Briefly, it makes disjoint clusters (categories) of the items. These clusters are such that items within a cluster tend to be co-consumed by a user and these co-consumption can be modeled using a single latent variable. Then, for each cluster a latent variable is added forming an LTM, and all these LTMs together form a forest as shown in Figure \ref{fig:forest}. We do not link up the categories in intermediate model and this is key for scalability.

Algorithm \ref{alg.buildislands} shows the procedure for building a flat forest model.  
The main procedure of learning a flat model is building one category.  This procedure is repeated until no more items are left and is outlined in Algorithm \ref{alg.oneisland}. We start by picking a random item from the set of all items $\mathcal{I}$ (line \ref{alg3.s1}). We could start by picking a pair of most similar items as in \cite{chen2017latent}, but that has a computational complexity of $\Omega(|\mathcal{I}|^2)$\footnote{It is $O(|\mathcal{I}|^2)$ for each category in the beginning and we have to learn $\mathcal{I}$ categories and this makes HLTA slow.}. Moreover, we found that the choice of picking a random item works well in practice and has a complexity of $O(1)$.

\begin{algorithm}[t]
\begin{description}
\item[Inputs:] $\mathcal{R}_1$, $\mu$, $\delta$, $\kappa$. \\
\item[Outputs:] A flat forest model $m_1$.
\end{description}

\begin{algorithmic}[1]
\State $\mathcal{I} \gets$ all items in $\mathcal{R}_1$, $m_1 \gets \emptyset$.
\While {$|\mathcal{I}| > 0$}
    \State $c \gets$ \sysname{OneCategory}($\mathcal{R}_1$, $\mathcal{I}$, $\delta$, $\mu$, $\kappa$);
    \State $m_1 \gets m_1 \cup \{c\}$;
    \State $\mathcal{I} \gets $ item in $\mathcal{R}_1$ but not in any $c \in m_1$;
\EndWhile
\State \Return $m_1$.
\end{algorithmic}
\caption{\sysname{LearnFlatForestModel}($\mathcal{R}_1$, $\delta$, $\mu$, $\kappa$)}
\label{alg.buildislands}
\end{algorithm}

We then find the most similar item, in terms of cosine similarity, to the items in the working set $\mathcal{S}$ (line \ref{alg3.s2}) and add it to the working set (line \ref{alg3.s3}). We use cosine similarity as a similarity metric for implicit feedback data instead of mutual information that is generally used for HLTA. Firstly, because it can scale up much better in terms of both memory and time requirements. Calculating mutual information has a time and memory complexity of $O(|\mathcal{U}||\mathcal{I}|^2)$ and $O(|\mathcal{I}|^2)$ respectively, where $\mathcal{U}$ is the set of all users. Moreover, it does not leverage the inherent sparsity present in the data. Cosine similarity, on the other hand, can be calculated much more efficiently using sparse data representations, as two items that were never co-consumed have zero similarity. This, also, drastically reduces the memory requirements as most item pairs have no cosine similarity. The time complexity of cosine similarity is $O(|\mathcal{U}|i_z^2)$, where $i_z$ is the average number of items consumed by a user and is generally small for sparse data. Secondly, since we are working with implicit feedback, the only reliable information we have is regarding item consumption. Cosine similarity for binary data effectively only relies on this information. This is unlike mutual information which assumes non-consumption to be equally informative as consumption.

An initial category LTM is then learned for the working set $\mathcal{S}$ of items (line \ref{alg3.firstlcm}). And then from lines \ref{alg3.pick-x} to \ref{alg3.add-x} we iteratively increase the number of items in this category until a stopping criteria is met i.e. either the maximum category size is reached (line \ref{alg3.maxisland}) or the category cannot be represented adequately by a single latent variable as determined by the UD-test (line \ref{alg3.compare}). We refer the reader to \cite{chen2012model} for UD-test and to \cite{Chen:2016:PEL:3016100.3016108} for details on the subroutines \sysname{Pem-Lcm} and \sysname{Pem-Ltm-2l}. The subroutine  \sysname{ProjectData} takes the users' consumption histories over all items and returns the consumption histories over the set of items in the second argument, and \sysname{LearnLCM} learns a latent class model (or LTM) given the set of items and the data over these items. Finally, for parameter tuning we run EM on the learned category $c$ for $\kappa$ steps. 

\begin{algorithm}[t]
\begin{description}
\item[Inputs:] $\mathcal{I}$, $\mathcal{R}_1$, $\mu$, $\delta$, $\kappa$. \\
\item[Outputs:] A category $c$.
\end{description}

\begin{algorithmic}[1]
\State \textbf{if} number of items $|\mathcal{I}| \leq 3$, $c \gets$ \sysname{LearnLCM}($\mathcal{R}_1$, $\mathcal{I}$), \textbf{return} $c$.\label{alg3.line1}
\State $\mathcal{S} \gets$ a random variable from $\mathcal{I}$ \label{alg3.s1}
\State {\scriptsize $X \gets \argmax_{A \in \mathcal{I} \setminus \mathcal{S}} CosineSim(A, \mathcal{S})$};\label{alg3.s2}
\State {$\mathcal{S} \gets \mathcal{S} \cup {X} $};\label{alg3.s3}
\State $\mathcal{I}_1 \gets  \mathcal{I} \setminus \mathcal{S}$;
\State $\mathcal{R}_2 \gets$ \sysname{ProjectData}($\mathcal{R}_1$, $\mathcal{S}$); \\
$c \gets$ \sysname{LearnLCM}($\mathcal{R}_2$, $\mathcal{S}$);\label{alg3.firstlcm}
\Loop
    \State {\scriptsize $X \gets \argmax_{A \in \mathcal{I}_1} CosineSim(A, \mathcal{S})$; \label{alg3.pick-x}
    \State $W \gets \argmax_{A \in S} CosineSim(A, X)$}\label{alg3.pick-w};
    \State $\mathcal{R}_2 \gets$ \sysname{ProjectData}$(\mathcal{R}_1, \mathcal{S} \cup \{X\})$, $\mathcal{V}_1 \gets \mathcal{V}_1 \setminus \{X\}$;
    \State $c_1 \gets$ \sysname{Pem-Lcm}$(c, \mathcal{S}, X, \mathcal{R}_2)$;\label{alg3.lcm}
    \State \textbf{if} $|\mathcal{I}_1| = 0$ \Return $c_1$.\label{alg3.return-lcm}
    \State $c_2 \gets$ \sysname{Pem-Ltm-2l}($c$, $\mathcal{S} \setminus \{W\}$, $\{W, X\}$, $\mathcal{R}_2$);\label{alg3.ltm}
    \If{$BIC(c_2|\mathcal{R}_2) - BIC(c_1|\mathcal{R}_2) > \delta$}\label{alg3.compare}
    	\State $c \gets$ Run EM on $c_2$ for $\kappa$ steps;
        \State \Return $c$ with $W$, $X$ and their parent removed.\label{alg3.return-island}
    \EndIf
    \If{$|\mathcal{S}| \geq \mu$}\label{alg3.maxisland}
    	\State $c \gets $Run EM on $c_1$ for $\kappa$ steps, \Return $c$.
    \EndIf
    \State $c \gets c_1$, $\mathcal{S} \gets \mathcal{S} \cup \{X\}$;\label{alg3.add-x}
\EndLoop

\end{algorithmic}
\caption{\sysname{OneCategory}($\mathcal{R}_1$, $\mathcal{I}$, $\delta$, $\mu$, $\kappa$)}\label{alg.oneisland}
\end{algorithm}

\subsection{Hard Assignment}
Once a flat model is learned (in line \ref{alg1:learnFlat} in Algorithm \ref{alg.top}), hard-assignment (line \ref{alg1.hardassign} in Algorithm \ref{alg.top}) is required to assign a particular state of each latent variable, of the newly learned flat model, to each user. This creates a new user by latent variable binary matrix $\mathcal{R}_1$. This matrix can now be treated as the feedback matrix, where the items are replaced by latent variables, and another layer of latent variables can be learned during \sysname{LearnFlatForestModel}.

The process of hard-assignment requires, for each user, calculating the posterior probability of each newly created latent variable. And then assigning the state with the higher posterior probabilty to this latent variable. In Previous works \cite{chen2012model,Chen:2016:PEL:3016100.3016108,chen2017latent} this posterior probability is calculated based on the current model $m$ which was created after the \sysname{StackModels} procedure. As a result, $m$ involves \emph{all} observed and latent variables. Moreover, $m$ is a \emph{tree} structured model. This means that inference requires message propagation over all latent and observed variables. Which is time consuming, $O(|\mathcal{I}|)$ for each user and $O(|\mathcal{U}||\mathcal{I}|)$ in total, and does not scale up to larger number of items. 

In our work, we do hard-assignment after learning each flat model and not after stacking the models. Therefore, inference only involves the variables in the current flat model and not all latent and observed variables. Moreover, since our flat model is a forest and not a tree, we don't require message propagation over all the variables in the flat model. Rather, the inference for each latent variable is done on its own tree that involves only its immediate children e.g. in Figure \ref{fig:forest} to get the posterior probability of $Z_{1223}$ for a user, we only need to do message passing over the nodes of four variables. Since, maximum number of nodes in a tree is upper bounded by $\mu$ this requires $O(1)$ time for each tree and $O(|\mathcal{U}|c)$ for all users, where $c$ is the number of latent variables (categories) at the current level.

\subsection{Complexity}
When building a category, HLTA generates roughly $O(I)$
intermediate models. Each Model estimates parameter in $O(1)$ time and data projection takes $O(u_z)$, where $u_z$ is the average number of users who consumed an item and is typically small. So total time for building categories in a layer is $O(|I|u_z)$. And the total time for learning a flat forest is $O(|\mathcal{I}|u_z + |\mathcal{U}|i_z^2)$. We can then add the time for hard-assignment so the total time complexity becomes $O(|\mathcal{U}|i_z^2 + |\mathcal{U}|c + |\mathcal{I}|u_z)$\footnote{Since, the number of levels in the model is typically small (less than seven) and the number of variable in making each subsequent level decrease, the overall complexity is the same.}. When the data is sparse, $i_z$ and $u_z$ are small and our time complexity is roughly linear in the size of the data. However, for vanilla HLTA it is $O(\mathcal{U}|\mathcal{I}|^2) + \Omega(|\mathcal{I}|^2)$. 

The running time comparison between HLTA and HLTA-Forest on three datasets\footnote{The details of the dataset are provided later.} is provided in Table \ref{tab:runtime}. The experiments were run on a machine with 4 x 10-core Intel Xeon 2.3GHz processors and 189GB RAM. HLTA could not complete a single layer on ML20M and xing datasets in 3 days, therefore their results are not provided. We see that HLTA-Forest is much faster and scales well to larger datasets. It is worth noting that since we use sparse representations for calculating cosine similarity in HLTA-Forest, when the cosine similarity matrix gets relatively dense, the sparse representations become slow (as seen for ML20M dataset). However, this can be easily mitigated by switching to dense representations if sufficient memory is available.
\begin{table}[]
\centering
\caption{Running time in seconds for HLTA and HLTA-Forest on three training datasets. HLTA could not complete a single layer in 3 days on the ML20M and xing datasets so their results are omitted. HLTA-Forest is much faster.}
\label{tab:runtime}
{\small
\begin{tabular}{lllll}
\hline
\textbf{lastfm}      & Cosine/MI & Flat Layer-1 & H.A. Layer-1 & Total Model \\
\hline
HLTA-Forest & 3.5        & 37.9         & 2.3                & 58.2        \\
HLTA        & 35.7       & 12549.2      & 2167.3             & 27460.6     \\
\hline
\textbf{ML20M}       &            &              &                    &             \\
\hline
HLTA-Forest & 2627.2     & 354.3        & 11.4s              & 3710.5      \\
\hline
\textbf{Xing}        &            &              &                    &             \\
\hline
HLTA-Forest & 231.6      & 4980.1       & 235.1              & 6596.1     \\
\hline
\end{tabular}
}
\end{table}

 \section{Browsing using HLTM}
The availability of item category information facilitates the users' browsing experience
by overcoming the information overload problem. The users can click the categories
that interest them the most, and can choose not to click the irrelevant categories, thereby
filtering them. Although, this form of filtering is done by the user themselves, but it
is made possible by the classification of items in well defined categories. 
 
 Once an HLTM is learned, we have a hierarchy of item categories where the leaves are represented by items and categories are represented by latent variables. However, the latent variable names don't have the semantic information to enable the user to browse through the item hierarchy. To make category names more informative, we replace the latent variable names with a few most representative items from that category. Then the user can easily know the kind of items represented in the category. 
 
\begin{figure*}

\begin{center}
\includegraphics[scale=0.6]{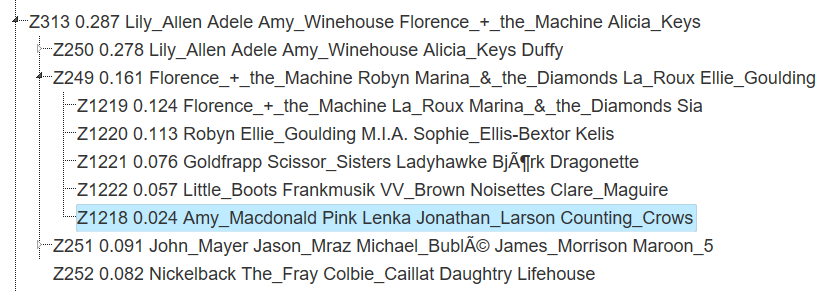}
\caption{Part of a sample browsing system where 5 representative items are used as category identifiers. Representative items are chosen based on MI. The category $Z_{1218}$ is highlighted. The users can click a category to see its sub-categories and browse through the item space. $Z_{1218}$ is pop-rock artists, while its ancestor category $Z_{313}$ is more general and represents artists who mix different genres with pop.  }
\label{fig:browse}
\end{center}
\end{figure*}

To select these few representative items for a category, we calculate mutual information between each item under this category and the latent variable for this category. Then we sort the items based on this mutual information and pick the top items as representatives for this category. For example, lets assume that we want to get the representative items for the first level category $Z_{1218}$. During the HLTM construction we have already calculated the conditional probabilities  $P(item | Z_{1218})$. Ideally, we would want our representatives items to have high probability of being 1 when the latent variable is also 1 (i.e. $state_1$) and low probability of being 1 when latent variable is 0. This is exactly what mutual information (MI) measures. Formally the MI $I(X; Y)$ between the two discrete variables $X$ and $Y$ is defined as follows:
 \begin{eqnarray}
I(X; Y) = \sum_{X, Y}P(X, Y) \log \frac{P(X, Y)}{P(X)P(Y)},
\label{eq.mi}
\end{eqnarray}
In Table \ref{tab:MI} we have sorted all the items in descending order w.r.t. their MI with $Z_{1218}$. We can then pick the top few e.g. 5 items as representative for this category. And replace the name $Z_{1218}$ with these top-5 item names as shown in Figure \ref{fig:browse} to make the categories interpret-able. We can see that the items under this category are primarily pop-rock therefore, the category represented by $Z_{1218}$ is pop-rock artists. In addition, MI is able to select suitable representatives as it selects artists that have a focus on both pop and rock and leaves out artists that are more towards the rock genre.

\begin{table}[]
\centering
\caption{The conditional probability $P(item=1 | Z_{1218})$ for both states of the latent variable is show. The items are sorted in descending order of MI with the latent variable. The items at the top have a high probability $P(item=1 |Z_{1218}=state_1$) and a low probability $P(item=1 | Z_{1218}=state_0$). Hence they have more MI with the category and can be taken as representatives. $Z_{1218}$ then represents pop-rock artists. }
\label{tab:MI}
\begin{tabular}{llll}
\hline
$\mathbf{Z_{1218}}$           & $\mathbf{state_0}$    & $\mathbf{state_1}$  & \textbf{Genre}   \\
\hline
Amy Macdonald   & 0.008 & 0.417 & Alternative rock, pop-rock\\
Pink            & 0.014 & 0.343 & R\&B, pop and rock\\
Lenka           & 0.003 & 0.137 & Pop, indie pop\\
Jonathan Larson & 0.005 & 0.093 & Rock, pop-rock \\
Counting Crows  & 0.005 & 0.076 & Rock, pop-rock\\
Ryan Cabrera    & 0.005 & 0.072 & Pop-rock\\
Acceptance      & 0.004 & 0.049 & Pop-rock, pop-punk\\
Noir D\`{e}sir      & 0.006 & 0.053 & Post-punk, alternative rock\\
Iggy Pop        & 0.009 & 0.043 & Rock, proto-punk\\
PJ Harvey       & 0.025 & 0.054 & Punk blues,art rock, indie rock\\
\hline
\end{tabular}
\end{table}

\subsection{Discussions}
Hierarchies exist naturally in many domains. For example, books can be categorized
by their topics; we can have a hierarchy in which books represent the items, and then a
collection of books is represented by a specific topic and a collection of specific topics can
be represented by a general topic and so on. Similar hierarchies exist in other domains
like songs, videos, and products in online shopping platforms.

Hierarchical organization is an effective way to quickly guide a user to the items of
interest. Platforms such as Amazon and Taobao have item taxonomies that are
created based on domain knowledge. They rely on sellers to place new items into
appropriate categories. The labels provided might be inaccurate and/or inconsistent
because they depend on human judgement. There are quality control polices and
housekeeping procedures. However, those do not completely eliminate the problem. In
addition, misclassifications can be deliberate. People might put their items under
irrelevant but popular categories to attract more traffic.

Automatic co-consumption based item hierarchy construction can help addressing the aforementioned problems.
It can also be useful when categories become too large. In that case, we might want to
extend the hierarchy by building sub-hierarchies. In addition, automatic hierarchy
construction can help small vendors who keep their own product inventories to create and
maintain item hierarchies.

HLTA obtains item hierarchies based solely on user behaviors and does not require
domain knowledge. As such, it is domain independent. On the other hand, because it relies on user
behaviors, HLTA can only be applied after sufficient transaction data are collected. Moreover, since these categories are based on item co-consumption, this approach is most useful in domains in which people co-consume semantically similar items. This is the case in most domains. Nevertheless, in cases where products like beer and diapers are being co-consumed, the semantic meaning of the category becomes fuzzy, as beer is a drink and diaper is not. However, although such categories are not ``true'' item categories, they might still be useful for users to find related items while browsing and might even result in users discovering serendipitous items.

\section{Category Aware Recommendation}
Each layer (level) of HLTA partitions all items into disjoint categories. We, can use these disjoint categories to make recommendations by setting the number of recommendations from each category proportional to the number of items
purchased before from that category. If a customer has bought many items from a category before, then the interest in that category is high and hence more items should be recommended.

Formally, For a given user $u$, we can choose a latent layer $l$ such that the purchased items are from a few categories $C_1, \dots , C_m$ on the layer, where $m$ is a small number (say 5). Let $n$ be the total number of
items consumed by the user and $n_i$ be the number of purchased items from category $C_i$.
Suppose were are to recommend $K$ items to the user. By category-aware recommendation
(CAR), we mean to set $r_{C_i}$, the number items to recommend from category $C_i$ as:

\begin{equation}
r_{C_i}=\frac{n_i}{n}K, \; s.t. \; n_i \geq \alpha,
\end{equation}

where, $\alpha$ is a tuneable threshold that define the minimum number of consumed items to be present in a category. This is done so that we make recommendations from a category if we have enough confidence on our estimation of the user's interest in that category. 

CAR can be combined with any base recommendation algorithms. It serves to ensure that the number of recommended
items from a category $C_i$ be roughly proportional to the ratio $n_i/n$ , which indicates how
much interest the user has in the category. Without CAR, the base algorithm might
recommend all items from a single category, which is clearly undesirable. The complete procedure is outlined in Algorithm \ref{alg.car}. 

After initialization we use depth first search on $m$ and get the set of all items in each category at level $l$ of $m$ (line \ref{alg.car.set}). Then we sort the base recommender's ranked list $B$ into each of these categories (line \ref{alg.car.base}) and get the counts $n_i$ (line \ref{alg.car.ni}). We then get the count of recommendations to be made from each category in line \ref{alg.car.count}. From line \ref{alg.car.cars} to \ref{alg.car.care} we make recommendation by first picking the top $r_{C_1}$ elements from the base recommender's list for the most consumed category $C_1$. These items are then deleted from the base recommender's lists (line \ref{alg.car.del}). Then we do the same for the second most consumed category and so on until we have made recommendations for the top $k$ categories. If our total recommendations are less than $K$, due to rounding or omitting categories, we follow the base recommender's list order and pick the top item (line \ref{alg.car.pick}). However, we want to make sure that it is not from a previously recommended category or a category from which the user has not consumed. If this is the case, we recommend it (line \ref{alg.car.new}). Otherwise, we move to the next item in the base recommender's list. 
\begin{algorithm}[t]
\begin{description}
\item[Inputs:] A HTLM $m$, category level $l$, base recommendation list $B$, set of items $\mathcal{R}_u$ consumed by user $u$,  $\alpha$, $K$. \\
\item[Outputs:] A top $K$ recommendation list $L$.
\end{description}

\begin{algorithmic}[1]
\State ${C_1 \dots C_m} \gets$ latent variable at level $l$ of $m$;
\State $n \gets |R_u|$, $n_i \gets 0 , i \in [1,m], B_{C_i}=\varnothing, C=\varnothing$;
\State $\forall C_i, i \in [1,m],$ get the set of items of this category $I_{C_i}$ from $m$; \label{alg.car.set}
\State $B_{C_i} \gets B_{C_i} \cup I \; s.t. \; I \in I_{C_i},\forall I \in B$;\label{alg.car.base}
\State $n_i \gets n_i+1 \; s.t. \; I \in I_{C_i}, \forall I \in \mathcal{R}_u$;\label{alg.car.ni}
\State sort $n_i, i \in [1,m], \; s.t. \; n_1>n_2 \dots > n_m$;
\For{$k =1 \dots m$}
\State \textbf{if} $n_k \geq \alpha$ \textbf{then} $r_{C_k}=\frac{n_k}{n}K$\label{alg.car.count}
\State \textbf{else break;}
\EndFor
\For{$j =1 \dots k$}\label{alg.car.cars}
\State $L \gets $ top $r_{C_j}$ elements from $B_{C_j}$;
\State $B \gets B\setminus L$ and $B_{C_j} \gets B_{C_j}\setminus L;$\label{alg.car.del}
\State $C \gets C \cup C_j;$
\EndFor\label{alg.car.care}
\While{$|L|< K$}
\State $I \gets$ top item from $B$\label{alg.car.pick}
\State  \textbf{if} $(I \in I_{C_i}) \wedge (I_{C_i} \notin C) \wedge (n_i \neq 0)$ \textbf{then} $L \gets I;$\label{alg.car.new}
\State $B \gets I \setminus B;$
\EndWhile
\State \textbf{return} $L$.
\end{algorithmic}
\caption{\sysname{CAR}($R_u$, $m$, $l$, $B$, $K$, $\alpha$)}\label{alg.car}
\end{algorithm}

Since the categories used in CAR are based on HLTM, the items within the categories can be viewed as items that a group of users co-consumed. By making recommendations from a category we are recommending items to a user that other users (who are also interested in this category) have co-consumed in the past. Moreover, if we don not recommend items from some categories than this is because users who had similar interest as our target user did not co-consume these items. Therefore, 
CAR rules out item spaces because of zero or weak co-occurrences. 

This also opens up door for using HLTA as a means for negative sampling in implicit feedback i.e., we can see the categories from which the user never consumed before and treat all those items as negative samples for the users and retrain the base recommender. This work is left as a future extension.

\subsection{Explaining Recommendations}
Using CAR on top of a base recommender introduces personalized structure into the list by recommending items from categories which the user has shown interest in. This, means that unlike the base recommenders, we explicitly know the reason why an item is in our recommendation list. Therefore, we can provide this reason as an explanation to the user. For example, user ID 32 in the lastfm dataset has consumed eight items from category $Z_{52}$. CAR makes a recommendation of {\tt Pixie Lott} from this category. We can then give the explanation ``Because of your interest in the category of artists like {\tt Jessie J.}, {\tt Nicki Minaj} and {\tt Adam Lambert} we recommend you to listen to {\tt Pixie Lott}''. Where, we take the explaining items from the representative items of the level-1 category, $Z_{1620}$, of {\tt Pixie Lott}, as they describe the recommended artist better. Note that the representative items of the category don't have to be items the user has consumed, therefore they too can be subtle/serendipitous recommendations. 

Providing explanations for recommendations is essential as it provides transparency to the user, builds the user's trust in the system, and improves the overall satisfaction. Since implicit feedback data does not have additional information about items, it is generally difficult to provide explanations using collaborative filtering methods. Although some specific models are capable of providing different types of explanations \cite{friedrich2011taxonomy,herlocker2000explaining} based on item similarities\footnote{Unlike CAR, if explanations are based on item similarity only, then we are limited to the items a user has consumed as reference items in the explanation.}, using similar users' preference or introducing new factors in the model etc. In this work we use our learned item hierarchies to give explanations. And, what makes this approach more attractive is that we are able to provide explanations for \emph{any} base recommender, by seeing the category of the recommended item, regardless of whether it was designed to extract such information or not.

\section{Related Work}

Item hierarchies can be built using a variety of methods. A well-known method is agglomerative hierarchical clustering (AHC) \cite{Duda:2000:PC:954544}, where one starts with each item as a separate cluster and merges the clusters recursively until a single cluster remains. AHC is distance-based.

There are also model-based methods for hierarchical clustering. The Bayesian approach \cite{neal2001defining,williams2000mcmc,griffiths2004hierarchical,roy2007learning} defines a prior over all trees and defines likelihood for data by assuming that data points within each cluster are independent draws from a distribution. A hierarchy is then obtained from the posterior via MCMC. Another approach \cite{Heller:2005:BHC:1102351.1102389,friedman2003pcluster} uses mixture models over sub-trees as the underlying probabilistic model. They
closely mirror the agglomerative clustering procedures, and the only difference is that the
decision to merge trees is based on posterior approximation.

All the methods mentioned above produce binary hierarchies, where each internal
node has only two children. Bayesian Rose Trees \cite{blundell2012bayesian} generalize Bayesian hierarchical
clustering \cite{Heller:2005:BHC:1102351.1102389} and obtain trees where an internal node potentially has multiple children.
It is also similar to agglomerative clustering, except that it decides whether to merge two
clusters based on the likelihood ratio between the merged clusters and the two separate
clusters. However, like other bayesian approaches mentioned, it does not scale up to collaborative filtering datasets as it has $O(|\mathcal{I}|^2)$ space and $O(|\mathcal{I}|^2 \log |\mathcal{I}|)$ time complexity.
\section{Experiments}
We used three publically available datasets for our experiments: Movielens20M (ML20M), last.fm, and xing dataset from RecSys'17 challange. Movielens20M\footnote{\url{https://grouplens.org/datasets/movielens/20m/}} is a dataset of users rating for movies, it was made implicit by ignoring the rating values. Last.fm is a dataset of user listening counts of music artists, the listening counts were ignored for our experiments. Finally, RecSys'17 challange xing\footnote{\url{www.recsyschallenge.com/2017/}} dataset contains events from a job portal. We treated all click,bookmark,reply and recruiter interest events as positive and retained users with greater than three events and items with greater than five events. The statistics of the datasets are shown in Table \ref{tab:stat}.

\begin{table}[]
\centering
\caption{Statistics of the datasets.}
\label{tab:stat}

\begin{tabular}{llll}
\hline
\textbf{last.fm} & \textbf{Users}  & \textbf{Items}  & \textbf{Sparsity} \\
\hline
train  & 1891   & 16048  & 99.74\%    \\
test   & 1880   & 5189   & 99.86\%    \\
\hline
\textbf{ML20M}  &        &        &          \\
\hline
train  & 118526 & 15046  & 99.05\%    \\
test   & 25561  & 25843  & 99.55\%    \\
\hline
\textbf{Xing}   &        &        &          \\
\hline
train  & 273500 & 120784 & 99.98\%    \\
test   & 241288 & 166884 & 99.99\%    \\
\hline
\end{tabular}

\end{table}

We split all datasets by a global time stamp such that all the instance before that time stamp go in the training set and the rest in the validation and test set respectively. This mimics real recommendation more closely. We used a 70-15-15 train, validation and test set split for our experiments. All the parameters were chosen based on the validation set.

We used three well known implicit feedback recommenders as our baseline recommenders: WRMF \cite{Hu:2008:CFI:1510528.1511352}, BPRMF \cite{Rendle:2009:BBP:1795114.1795167} and WeightedUserKNN and WeightedItemKNN. The MyMediaLite\cite{Gantner:2011:MFR:2043932.2043989} implementation was used for all baseline recommenders.

\subsection{Evaluation Metrics} 
We use two accuracy metrics, precision@$N$ ($P@N$) and recall@$N$ ($R@N$) and two diversity and personalization metrics. For diversity we use the global diversity ($D@N$) \cite{Adomavicius:2012:IAR:2197072.2197127} which is the total number of unique items recommended by the system. And for personalization we measure inter-user diversity \cite{zhou2010solving} that measures the individuality of each user's recommendation list by : $H@N\equiv1 - \frac{q_{ij}(N)}{N}$, where $q_{ij}(N)$ is the number of common items in the top-$N$ lists of users $i$ and $j$.

\subsection{Accuracy and Diversity}
CAR ensures that each of the user's dominant tastes are catered for in the recommendation list. This leads to the list for an individual user to be diverse. Moreover, since different users are interested in different categories, recommending using CAR results in personalized recommendations. Therefore, the overall system diversity also increases. In Table \ref{tab:acc} we show the accuracy and diversity results\footnote{Due to the large number of test users in xing, validation of parameters is time consuming, therefore its accuracy-diversity results are not reported.}. We can see that as expected, base recommenders with CAR are more personalized and diverse with little compromise in accuracy. These results also suggest that the item categories learned by HLTA-Forest are meaningful and group related items together.

With CAR we also ensure that sudden changes in a user's interest won't take immediate effect. It is only after a user consumes a minimum amount of items (governed by $\alpha$) from a category that CAR will accept this category for making recommendations. Moreover, as a user sees most of his interests represented in the recommendation list, this encourages the user not to be sucked in a ``filter bubble''\cite{Nguyen:2014:EFB:2566486.2568012}.

\begin{figure}

\begin{center}
\includegraphics[scale=0.35]{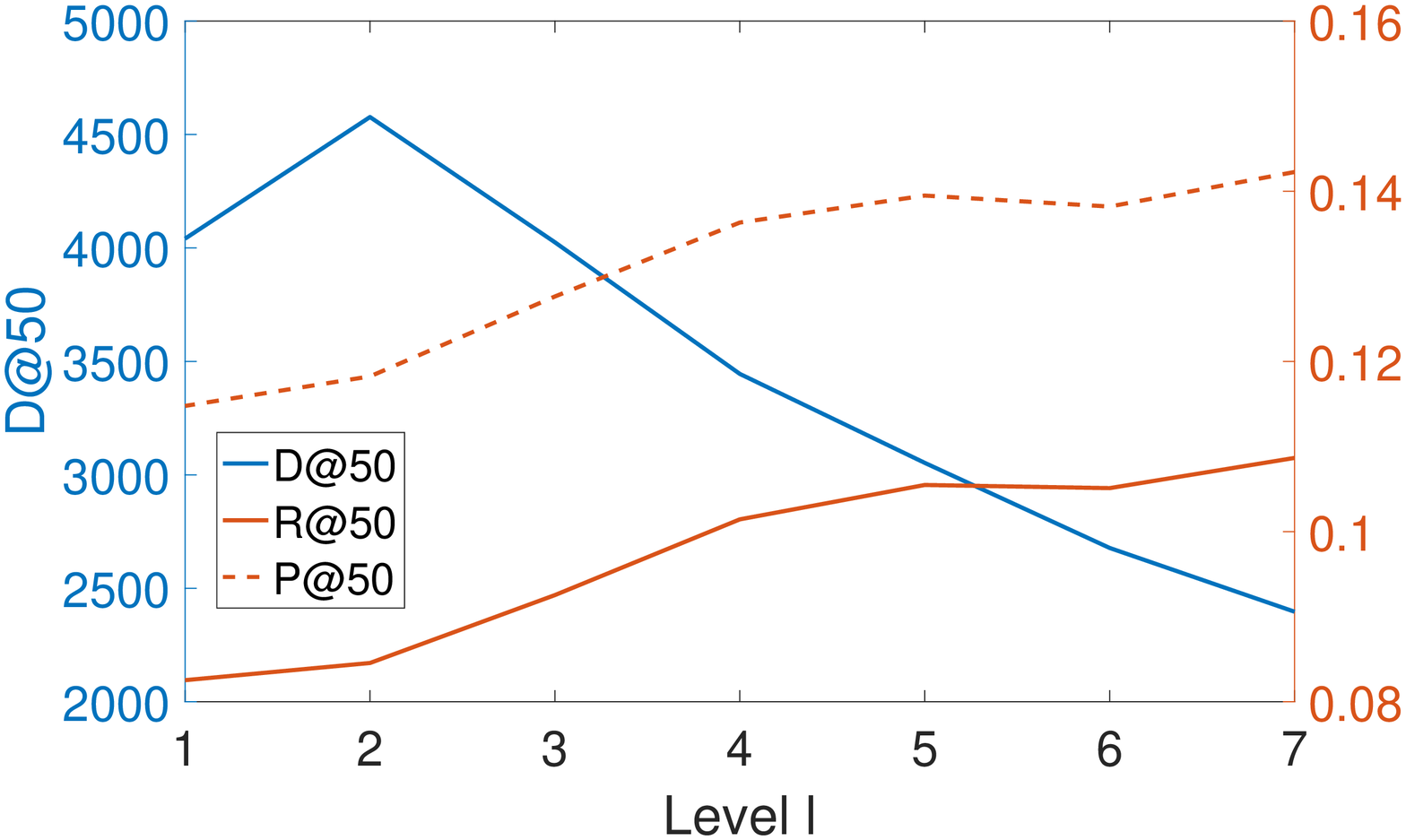}
\caption{Plot of P@50/R@50 on the right axis and D@50 on the left axis against the level $l$ of the HLTM for ML20M. As we go down the hierarchy the categories become more specific and diversity increases. However, the accuracy decreases as the number of categories increase and there is not enough data in each category to reliably ascertain the user's preference.}
\label{fig:acc-div}
\end{center}
\end{figure}

In Figure \ref{fig:acc-div} we show the accuracy-diversity trade off with $l$. As we go down the hierarchy the number of categories increase and become more specific. This results in increased diversity as we now force the model to recommend from more and specific categories. This also results in decreased accuracy as the user history, which is already limited, is not sufficient to estimate the preference for these categories.

\begin{table}[]
\centering
\caption{We can see that for all cases, CAR improves the diversity and personalization of the recommendation list without much loss in accuracy. The selected paramters for CAR are also shown. WeightedUserKNN could not run on ML20M due to high memory complexity. }
\label{tab:acc}
{
\begin{tabular}{lllll}
\hline
\textbf{ML20M}                       & \textbf{R@50} &\textbf{ P@50} & \textbf{D@50} & \textbf{H@50}						\\
\hline
WRMF                        & 0.12484   & 0.15688 & 2430                                            & 0.90939                                 \\
CAR\_WRMF\scriptsize{($l=5,\alpha=5$)}   & 0.12227   & 0.15445 & 2781                                            & 0.91179                                 \\
WIKNN                       & 0.10866   & 0.14226 & 2397                                            & 0.8076                                  \\
CAR-WIKNN \scriptsize{($l=5,\alpha=5$)}  & 0.10548   & 0.13949 & 3053                                            & 0.81327                                 \\
BPRMF                       & 0.11593   & 0.14489 & 2330                                            & 0.74558                                 \\
CAR-BPRMF \scriptsize{($l=5,\alpha=5$})  & 0.11022   & 0.13995 & 2544                                            & 0.76398                                 \\
\hline
\textbf{last.fm}                      &           &         &                                                 &                                         \\
\hline
WRMF                        & 0.42952   & 0.05802 & 1194                                            & 0.89636                                 \\
CAR\_WRMF \scriptsize{($l=5,\alpha=10$)} & 0.41902   & 0.05682 & 1432                                            & 0.90037                                 \\
WIKNN                       & 0.46156   & 0.0611  & 3118                                            & 0.8795                                  \\
CAR-WIKNN \scriptsize{($l=5,\alpha=10$)} & 0.4469    & 0.05946 & 3173                                            & 0.88494                                 \\
BPRMF                       & 0.28213   & 0.03909 & 1214                                            & 0.73861                                 \\
CAR-BPRMF \scriptsize{($l=5,\alpha=10$)} & 0.28032   & 0.03887 & 1291                                            & 0.75092                                 \\
WUKNN                       & 0.34187   & 0.04657 & 852                                             & 0.67955                                 \\
CAR-WUKNN \scriptsize{($l=5,\alpha=10$)} & 0.33639   & 0.04596 & 1064                                            & 0.70054                                
\\
\hline
\end{tabular}
}
\end{table}

\section{Conclusion}
We have presented a fast and scale-able method, HLTA-Forest, for learning hierarchical item categories from implicit feedback data. Since it is a probabilistic structure, we can assign informative category representatives, without any additional information source, to make browsing easier. We argue that it can also be useful for online vendors to create and maintain product hierarchies. We have shown that by making use of the learned item categories we can make existing recommenders diverse and personalized without much compromise in accuracy. Both of these are desirable properties for a recommender system. Moreover, this framework also allows us to provide explanations for any existing recommender. As a future study it would be interesting to use the item categories for negative sampling in implicit feedback.
\section*{Acknowledgment}
Research on this article was supported by Hong Kong Research Grants Council under grant 16202118.

\bibliographystyle{ACM-Reference-Format}
\bibliography{CoF}

\end{document}